\documentclass [12pt,a4paper]{article}

\usepackage[cp1251]{inputenc}
\usepackage[english]{babel}

\usepackage[T2A]{fontenc}
\usepackage{amsmath,amssymb,amsfonts,amsthm}
\usepackage{color}
\usepackage[pdftex]{graphicx}
\usepackage[pdftex,colorlinks,linkcolor=blue, citecolor=blue]{hyperref}

\title{\vspace{-4ex}\large \textbf{Tidal Disruption of Stars by Supermassive Black Holes\\ and Naked Singularities with Scalar Hair}}

\author{Eduardo Andre\:\!$^{1,2}$ and Alexander Tsirulev\:\!$^{2}$\:\!\thanks{tsirulev.an@tversu.ru} \vspace{1ex}\\
$^{1}$\:\!{\small \textit{Faculty of Sciences, Agostinho Neto University, Avenida 4 de Fevereiro 7, Luanda, Angola}}\\
\hspace{-3em}$^{2}$\:\!{\small \textit{Faculty of Mathematics, Tver State University, 35 Sadovyi, Tver, Russia, 170002}}}

\date{}

\begin{document}

\maketitle
 
\renewcommand{\abstractname}{}\abstractname
\vspace{-6ex}
\begin{abstract}
In this paper we study tidal forces near strongly gravitating objects at the centers of galaxies. In our approach, dark matter surrounding the centers of galaxies is modeled by a nonlinear scalar field. We focus on static, asymptotically flat, spherically symmetric black holes and naked singularities supported by a real self-gravitating scalar field minimally coupled to gravity. We consider the influence of dark matter on the tidal forces caused by the central objects and some features of the motion of matter near these configurations. It turns out, first, that the event horizon radius of a scalar field black hole can be arbitrarily small as well as the radius of the innermost stable circular orbit, and tidal forces near this orbit can be very large. Second, we show that if strongly gravitating objects in the centers of galaxies are naked singularities, then bright flares (tidal disruption events) in the nuclei of galaxies can be explained in different way. In the vicinity of a scalar field naked singularity, baryonic matter is concentrated in the gravitational potential well forming a "gray shell" consisting of cold compressed matter. In this case, the flares are caused by the collisions of stars with this shell.
\end{abstract}

\section{Introduction}

Bright flares in the nuclei of inactive galaxies are presently interpreted as tidal disruption events (TDEs), which are induced by strongly gravitating objects located at the centers of galaxies~\cite{Carter1983, Evans1989, Rees1998, Komossa2015}. These events open a new channel for exploring accretion disks and the dynamics of galactic nuclei. There are also reasonable expectations that future observations of TDEs will clarify the nature of the central objects~\cite{Bellm2019, Ivezic2021, Gezari2021}. It is currently believed that supermassive black holes (BHs) are the most likely candidates for the role of the central objects. In fact, however, there is still some uncertainty in our knowledge of the structure of spacetime in the vicinity of galactic centers. Other possibilities that are considered in the literature~\cite{Johannsen2016} include, in particular, wormholes~\cite{Kardashev2007, Dai2019, Potashov2020, Banerjee2021}, naked singularities (NSs)~\cite{Joshi2014, Goel2015, Shaikh2018, Potashov2019} or a dense core with the same outer geometry~\cite{Becerra-Vergara2021}, and boson stars~\cite{Kunz2021, Grould2017}. In this paper, we focus on static, asymptotically flat, spherically symmetric BHs and NSs supported by a real self-gravitating scalar field minimally coupled to gravity, partly because these configurations can be treated in one and the same manner, and largely because the behavior of matter in the central regions of scalar field NSs provides an alternative explanation of TDEs.

Astronomical observations can be interpreted only within the framework of a specific mathematical model. We assume that any physically relevant model of a galactic center should have the following three important features. First, one should not think of the central objects in galaxies as being in vacuum, because dark matter (DM) surrounding the centers of galaxies cannot be ignored. In our approach, DM is \textit{modelled} by a nonlinear scalar field~\cite{Matos2004, UrenaLopez2019, Konoplya2022} representing a kind of anisotropic fluid, so that it does not matter whether scalar fields exists in nature. The microscopic description of scalar field DM at the particle level encounters some difficulties (since we do not have an efficient scheme for quantization of a nonlinear scalar field), but the anisotropy of energy-momentum suggests that the nature of the corresponding particles is beyond the Standard Model. Note in this connection that the nature of DM is one of the unsolved problems of physics. The most common phenomenological model of cold DM assumes that DM consists of weakly interacting massive particles, which are not currently detected. The proposed equations of state for cold DM are isotropic, of the form $P=f(\rho)$, so that these two models are quite different, but the scalar field model of dark matter better explains some observational data~\cite{Lee1996, Brax2020, Shapiro2022}, in particular, resolves the cusp-core problem. Second, we expect that the flares caused by tidal destroys occur at distances less than a few hundred radii of the innermost stable circular orbit (ISCO). In such a region, the distribution of DM and, consequently, the spacetime geometry are unknown (the Schwarzschild metric is inappropriate at these distances). Therefore, it is necessary to ensure rigorous consistency of the model with general relativity; in particular, all physically meaningful functions must obey the Einstein field equations. In order to obtain the required structure of spacetime geometry in the center of a galaxy, we have the considerable degree of freedom in the choice of a scalar field self-interaction potential. On the other hand, current observations of the Galactic Center S-star cluster show that at a distances of about a thousand Schwarzschild radii we deal with small perturbations of Newtonian gravity~\cite{Borka2021, Abuter2022}. Thus, the third natural requirement for a correct model of a galactic center is that the internal exact solution smoothly fits the external solution, which, in turn, must fit the rotation curve data. Note that in addition to the models listed above, tidal forces are also intensively studied in other spacetime geometries~\cite{Crispino2016, Shahzad2017, Lima2020, Vandeev2021}.

Our main goal in this paper is to compare tidal forces near the center of a BH and near the center of a NS of the same mass in the scalar field DM model. These types of scalar configurations demonstrate a very significant difference in the features of tidal forces in the central region, both from each other and from the Schwarzschild vacuum black hole. In particular, the radii of the event horizon and the corresponding ISCO of a scalar field BH are always smaller (they can be very small and even arbitrarily close to zero) than those of the Schwarzschild solution with the same mass. On the other hand, a scalar field NS should apparently be surrounded by a spherical shell consisting of compressed gas or solid matter, so that the nature of the flares should differs from that near a supermassive BH. At the same time, as follows from numerical experiments with physically relevant solutions, the behavior of the metric functions for these three types of configurations of the same mass practically does not differ at distances exceeding ten Schwarzschild radii.

The paper is organised as follows. Section~\ref{Sec2} contains the necessary mathematical background for static, spherically symmetric scalar field configurations. In particular, we describe the so called method of restored potential and write out a general solution of the static Einstein-Klein-Gordon equations in the form of quadratures. In Section~\ref{Sec3} we consider the spacetime metrics in the vicinity of BHs and NSs with scalar hair and compare them with the Schwarzschild metric. Using a family of exact solutions as an illustrative example, we discuss the corresponding spacetime geometries, as well as the shape of effective potentials and possible types of orbits of massive test particles. Section~\ref{Sec4} is devoted to computations of tidal forces for the scalar field configurations under consideration. Within the framework of the accepted model, we also discuss the main question of how to distinguish BHs from NSs using available and future observational data on TDEs. We adopt the definition $R^i_{jkl}=\partial_k\Gamma^i_{jl}-\ldots$ for the curvature and the signature $(+,-,-,-)$ for the metric. Throughout the paper, we use the geometric system of units with $G=c=1$.

\section{Action, Field Equations, and Quadratures} \label{Sec2}

Our model of BHs and NSs with scalar hair is based on the Hilbert--Einstein action extended by a minimally coupled real scalar field. The action has the form
\begin{equation}\label{action}
    S=\frac{1}{8\pi}\int\left(-\frac{1}{2}R+ \langle d\phi,d\phi\rangle-2V(\phi)\right) \sqrt[]{|g|}\,d^{\,4}x\,,
\end{equation}
where $R$ is the scalar curvature, $V(\phi)$ is a self-interaction potential of a nonlinear scalar field $\phi$, and the angle brackets denote the pointwise scalar product induced by the spacetime metric $g$.

In general, the spacetime geometry at the centers of galaxies should be considered as stationary and axisymmetric, but we will ignore the effects of rotation. We justify this approach by the following two arguments: (i) many of the central objects, including Sgr\,A*, have very small spin~\cite{Fragione2020, Reynolds2022}; (ii) sometimes a simplified but exact model allows us to find new effects that turn out to be hidden in a more realistic but approximately-phenomenological description. Thus, we will dealing with a static, spherically symmetric, asymptotically flat spacetime, and therefore it is convenient to write the corresponding metric in the Schwarzschild-like coordinates as
\begin{equation}\label{metric}
    ds^2=A dt^2- \frac{\,dr^2}{f}- r^{2}(d\theta^2+\sin^{2}\!\theta\, d\varphi^2), \quad
    A=\mathrm{e}^{2F}\!f.
\end{equation}
The field $\phi$ and the metric functions $A, F, f$ depend only on the radial coordinate $r$ and satisfy the asymptotic conditions
\begin{equation}\label{cond0}
\phi= O\!\left(r^{-1/2-\textstyle{\alpha}}\right)\!, \quad \mathrm{e}^{F}=1+o\!\left(r^{-1}\right)\!, \quad A=1-\frac{2M}{r}+ o\!\left(r^{-1}\right)\!, \quad r\rightarrow\infty\!\:,
\end{equation}
where $\alpha>0$ and $M$ is the Schwarzschild mass. Varying the action~(\ref{action}) with respect to $g$ and $\phi$, one obtains the Einstein-Klein-Gordon equations
\begin{equation}\label{EKG1}
-\frac{f'}{r}-\frac{f-1}{r^2}={\phi'}^2 f + 2V\,,
\end{equation}
\begin{equation}\label{EKG2}
\frac{f}{r}\left(\!2F'+\frac{f'}{f}\right)+ \frac{f-1}{r^2}=
{\phi'}^2 f - 2V\,,
\end{equation}
\begin{equation}\label{EKG3}
-f\phi''-\frac{\phi'}{2}f'-\phi' f\left(\!F' +
\frac{1}{2}\frac{f'}{f}+\frac{2}{r}\right)+ \frac{dV}{d\phi}=0\,,
\end{equation}
where a prime denotes differentiation with respect to $r$.

Various solutions to these equations can be found by the "restored potential method" or, in alternative terminology, the "inverse problem method for a self-interacting scalar field minimally coupled to gravity"~\cite{Lechtenfeld1995, BronnikovShikin2002, Tchemarina2009, Azreg2010}. We will use below a variant of this method in the form of the quadratures~\cite{Solovyev2012}
\begin{equation}\label{F-xi}
F(r)=-\!\int_{r}^{\,\infty}\!\! {\phi'}^{2}rdr\,,\quad \xi(r)=r+\int_{r}^{\,\infty}\!\!
\left(1-\mathrm{e}^F\right)\!dr\,,
\end{equation}
\begin{equation}\label{A-f}
A(r)=2r^{2}\!\!\int_{r}^{\,\infty}\! \frac{\,\xi-3M}{\,r^4}\,\mathrm{e}^{F}dr\,, \quad f(r)=\mathrm{e}^{-2F}A\,,
\end{equation}
\begin{equation}\label{V}
\widetilde{V}(r)=\frac{1}{2r^2}\!\left(1-3f+ r^2{\phi'}^{2}\!f+ 2\,\mathrm{e}^{-F}\,\frac{\,\xi-3M}{r}\right),
\end{equation}
where $\widetilde{V}(r)=V(\phi(r))$ and it is required that one of the function $\phi\!\:,\!\: F$ or $\xi$ is given. Thus, the quadratures~(\ref{F-xi}) and~(\ref{A-f}) determine the metric functions, while the formula~(\ref{V}) determines the self-interaction potential as a function of the radial coordinate. Next, assuming that $\phi(r)$ is piecewise monotone, one can restore the potential $V(\phi)$. It can be seen directly from the quadratures that for all $r>0$
\begin{equation}\label{cond1}
F\leq0\!\:,  \quad \mathrm{e}^F\leq1\!\:,  \quad \xi>0\!\:, \quad \xi'=\mathrm{e}^F>0, \quad  \xi''= r{\phi'}^2\mathrm{e}^F\geq0\!\:.
\end{equation}
The boundary conditions for $\xi$ are
\begin{equation}\label{cond2}
\xi\!=\xi(0)+\alpha{}r+O\big(r^2\big)\;\, (0\leq\alpha\leq1)\!\:, \;\; r\rightarrow0\!\:,
\qquad
\xi\!=r+o(1)\!\:, \;\; r\rightarrow\infty,
\end{equation}
and $\xi(0)>0$ if $\phi'$ is not identically zero. Indeed, in accordance with the conditions~(\ref{cond1}), $\xi$ is a strictly monotonically increasing and strictly convex downwards function (of class $\mathcal{C}^2$ at least); this implies $\xi(r)\geq{r}$ for $0\leq{r}\leq\infty$, $\xi(0)\geq{0}$, and $\alpha\geq{0}$. Note that these conditions for $\xi$ are true even if $\phi$ is singular at the origin and $\mathrm{e}^{F(r)}\rightarrow0$ as $r\rightarrow0$. On the other hand, the behavior at infinity follows directly from~(\ref{cond0}).

In what follows, our strategy will be to start with a function $\xi(r)$ satisfying the conditions~(\ref{cond2}), and then to compute the other functions in accordance with the scheme
\begin{equation}\label{}
\xi \rightarrow \mathrm{e}^{F}\!\!= \xi' \rightarrow A \rightarrow f\!\:,
\qquad
\mathrm{e}^{F}\! \rightarrow F \rightarrow \phi'\!=\!\sqrt{F'/r} \rightarrow \phi \rightarrow \widetilde{V}\!\!\;(r) \rightarrow V\!\!\;(\phi)\!\:.
\end{equation}
Note that the substitution of the quadratures into equations~(\ref{EKG1})\,--\,(\ref{EKG3}) reduces the latters to identities for any choice of the function $\xi$.

\section{Black Holes and Naked Singularities}
\label{Sec3}

It follows from~(\ref{EKG2}) that the type of solution is determined only by the parameters $\xi(0)$ and $M$. BHs are obtained if $\xi(0)<3M$, since the integrand in~(\ref{A-f}) becomes negative near the origin. The parameter domain $\xi(0)>3M$ corresponds to NSs. A solution with $\xi(0)=3M$ is either a BH, a NS, or a regular solution, depending on the behavior of $F(r)$ (or, equivalently, $\phi(r)$) near the origin; the detailed analysis of the possible types of solutions to equations~(\ref{EKG1})\,--\,(\ref{EKG3}) is given in~\cite{Solovyev2012}. Leaving aside solutions with $\xi(0)=3M$, which in each case require fine tuning of the distribution of the scalar field $\phi(r)$ and the mass $M$, we will focus on the most general types of BHs and NSs. First of all, we will consider the principal difference in the form of $A(r)$ in these two types of configurations, since this metric function determines the parameters of orbits of test particles, the radius of ISCO (if it exists), the magnitude of tidal forces, and, ultimately, the intensity of tidal disruption flares.

The main feature that distinguishes BHs from scalar NSs is the following. In a scalar field BH spacetime defined by the quadratures (\ref{F-xi})\,--\,(\ref{V}) and the conditions~(\ref{cond0}), (\ref{cond1}), and (\ref{cond2}), the metric function $A(r)$ is a strictly increasing function outside the event horizon $r=r_h$~\cite{Potashov2019}. On the other hand, in a scalar field NS spacetime, $A(r)$ obviously decreases from infinity (at $r=0$) to its minimum value $A_m=A(r_m)$ (at some radius $r=r_m$) and then increases to the value $A_\infty=1$. In the asymptotic region, $A(r)$ in both cases is very close to the Schwarzschild BH solution of the same mass; note that for the Schwarzschild NS (with $M<0$), $A(r)$ is strictly decreasing function and $A(r)>1$ for all $r>0$, so that the asymptotic condition~(\ref{cond0}) does not hold.

The points $r_h$ and $r_m$ have a key role in our analysis, because they determine the basic properties of the effective potentials of test particles. First, an obvious key property of scalar field BHs is that $r_h<2M$ and $r_h\rightarrow0$ as $\xi(0)\rightarrow3M+0$; moreover, numerical experiments show that $r_{\!\scriptscriptstyle{I\!\!\;SCO}}<3r_h<6M$ (for a Schwarzschild BH $r_{\!\scriptscriptstyle{I\!\!\;SCO}}=6M$). This means that the curvature of spacetime and, consequently, tidal forces in the vicinity of the ISCO can be arbitrarily large, so that the upper limit of masses of supermassive BHs, for which the disruption of a main sequence star is possible, may be much greater than $\sim10^7M_\odot$ (the upper limit for the vacuum case). Second, in the case of a scalar field NS, the existence of a minimum of $A(r)$ implies that there is a minimum of the effective potential of a massive test particle at rest (that is, with zero specific angular momentum and zero radial velocity). Therefore, ordinary baryonic matter will lose energy through radiation and then, over time, will be concentrated in a spherical shell around the sphere of radius $r_m$.

Note that the bulge size in galaxies depends on the type and size of the galaxy, but is typically of order $10^{21}-10^{22}$\,cm. According to modern estimates, the DM mass in the bulge is from one-fifth to one-third of its total mass. For the Milky Way, the DM mass is $\sim10^{11}M_\odot$, while the BH mass is $\approx4.3\cdot10^6M_\odot\approx6\cdot10^{11}$cm. In the region of a few hundred Schwarzschild radii, baryonic matter is captured by a BH (or the potential well of a NS) and does not significantly affect the spacetime geometry. In fact, the baryonic matter in the central region is concentrated in the accretion disk, the mass of which is relatively small; e.g., for Sgr A*, it is less than $10^{-4}M_\odot$~\cite{Blandford2019}.

To illustrate this crucial difference, we next consider a one parameter family of solutions that includes both BHs and NSs, and highlight the common features in our analysis. For simplicity of computations and obtaining solutions in a fully analytical form, we define a fully analytic function $\xi$ in the interval $(0,\infty)$ by the formula
\begin{equation}\label{}
\xi(r)= \sqrt{r^2+ar+a^2}-\frac{a}{2}\,,
\end{equation}
so that
\begin{equation}\label{}
\mathrm{e}^{F(r)}= \frac{a+2r}{2\sqrt{r^2+ar+a^2}}\,,\qquad \phi'(r)= \frac{a}{\sqrt{(2r/3)(a+2r)(r^2+ar+a^2)}}\,,
\end{equation}
where the parameter $a>0$ determines the concentration/fuzziness of the scalar field near the center ($a=0$ corresponds to flat spacetime). It is not difficult to show that these functions ensures the correct fulfillment of the conditions~(\ref{cond1}) and asymptotics~(\ref{cond2}).
We do not need an explicit expression for the field function (in terms of the incomplete elliptic integral of the first kind), and we note only that it decreases at infinity as $r^{-1}$ and diverges logarithmically at $r=0$.

The direct integration in~(\ref{A-f}) yields  the metric function
\begin{multline}\label{A}
A(r)= 1+\frac{a}{3r}- \sqrt{r^2+ar+a^2} \left(1+\frac{7r}{4a}- \frac{37r^2}{8a^2}\right)\!\frac{a+6M}{6ar}\\
-3\ln\left[5+\frac{8a}{r}+\frac{8a^2}{r^2}+ \left(4+\frac{8a}{r}\right)\!\!\left(1+\frac{a}{r}+ \frac{a^2}{r^2}\right)^{\!\!1/2}\right] \frac{(a+6M)r^2}{64a^3}\quad\\
+\left(3\ln\!\!\:3-\frac{74}{3}\right) \frac{(a+6M)r^2}{32a^3}\,.
\end{multline}
Its behavior at infinity is
\begin{equation}\label{}
A(r)= 1-\frac{2M}{r}+ \frac{3a^3+18Ma^2}{40r^3} + O(r^{-4})\,, \;\; r\rightarrow\infty\,,
\end{equation}
and near the center
\begin{equation}\label{}
A(r)= \frac{a-6M}{6r}+ \frac{5a-18M}{8a}+ + \frac{9a+54M}{16a^2}\:\!r+ O(r^2)\,, \;\;r\rightarrow0\!\:.
\end{equation}
Then it can be seen from~(\ref{A}) that the coefficient of $1/r$ is negative (positive) if $a<6M$ (respectively $a>6M$) and the corresponding solutions are BHs (respectively NSs). In the case of fine tuning, $a=6M$, we obtain the regular configuration which does not have the photon orbit, the ISCO, the event horizon, and the Kretchmann invariant, $R_{\alpha\beta\gamma\delta} R^{\alpha\beta\gamma\delta}/4$, is bounded everywhere.

\begin{figure}[]
\begin{center}
\includegraphics [width=0.483\textwidth]{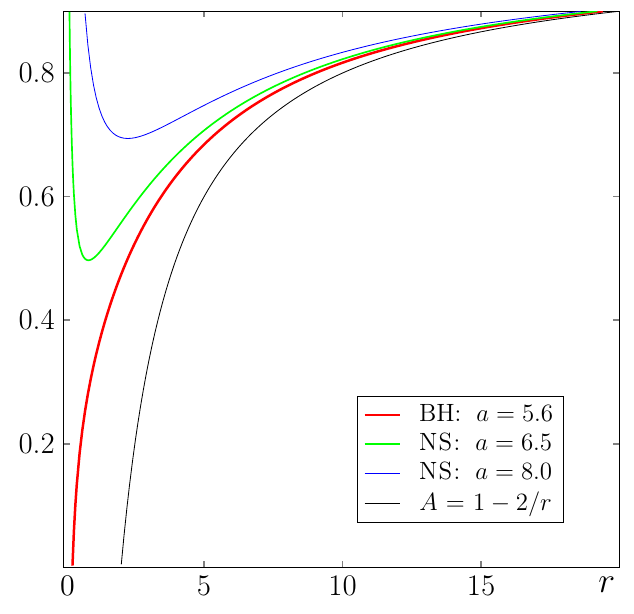}
\includegraphics [width=0.483\textwidth]{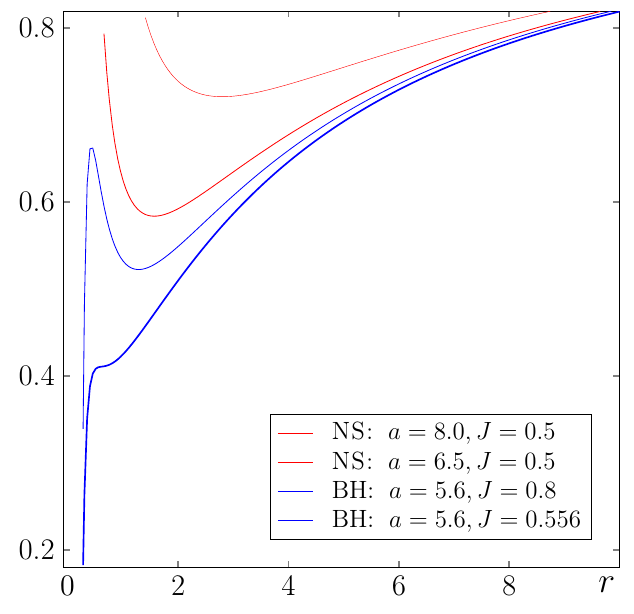}
\end{center}
\caption{Left panel: the metric functions $A(r)$ given by the expression~(\ref{A}) for various values of the parameter~$a$ and the Schwarzschild solution with the same mass ($M=1$). Note that the event horizon radius of the BH with~$a=5.6$ is $r_h\approx0.31$ (about 15\%
of the Schwarzschild radius). Right panel: the corresponding effective potentials of freely falling  massive particles determined by the expression~(\ref{U}). The value~$J=0.556$ for the BH with~$a=5.6$ corresponds to the ISCO of the radius~$r_{\!\scriptscriptstyle{I\!\!\;SCO}} \approx 0.74<3r_h\approx0.93$.}
\label{fig1}
\end{figure}

We are interested in the behaviour of $A(r)$ in the inner region, roughly speaking, with radii $r<100M$, since it is assumed that the flares associated with the tidal disruption of stars occur in such a region of galactic centers. In the outer region, one can continue this function in another way than required by the expression~(\ref{A}). If a whole galaxy were under consideration, it would be necessary to agree the outer geometry with the current observations of rotation curves in galactic halos. In doing so, $A(r)$ should be specified as a piecewise analytic function at least of class $\mathcal{C}^2$ at the point $r=100M$. Actually, any other reasonable choice of the matching point is appropriate. Moreover, the left panel in Fig.\,\ref{fig1} shows that the function $A(r)$ of our model practically coincides with the corresponding Schwarzschild metric function in the region $r>20M$ for both scalar field BHs and NSs. It is true for the metric function $\mathrm{e}^{F(r)}$ which is very close to unity for $r>20M$. In particular, for the physically appropriate values $0<a<10$, $A(r)$ is practically independent of $a$.

We fix the unit of length by setting $M=1$ in what follows. The three first integrals of the geodesic equation and the effective potential $\mathcal{V}_{\!e}$ of a freely falling  massive particle have the form
\begin{equation}\label{int-motion}
\frac{dt}{ds}=\frac{E}{A(r)}\,,\quad
\frac{d\varphi}{ds}=\frac{J}{\,r^2}\,,\quad
\left(\frac{dr}{ds}\right)^{\!\!2}=  \mathrm{e}^{-2F(r)} \left(E^2- \mathcal{V}_{\!e}(r)\right),
\end{equation}
\begin{equation}\label{Ve}
\mathcal{V}_{\!e}(r)= A(r)\left(1+\frac{J^2}{r^2}\right),
\end{equation}
where $E$ and $J$ are, respectively, the specific energy and the specific angular momentum of the particle. The behavior of the curves in Fig.\,\ref{fig1} illustrates the main features of scalar field configurations, which have been discussed above on the base of the general analysis of the quadrature formulae (\ref{F-xi})\,--\,(\ref{V}) and the asymptotic conditions~(\ref{cond2}). First, the BH with $a=5.6$ has the event horizon radius~$r_h\approx0.31$, which is almost seven times less than the Schwarzschild radius. Moreover, the corresponding ISCO (the right panel in Fig.\,\ref{fig1}) has the radius~$r_{\!\scriptscriptstyle{I\!\!\;SCO}} \approx0.74$. Note also that the radius $r_{ph}$ of photon orbit for the metric functions~(\ref{F-xi}) and~(\ref{A-f}) is determined by the equality~$\xi=3M$; it can be shown that always $r_{ph}<3M$ for scalar field BHs. A qualitative explain of an arbitrarily small size of the event horizon is that a static configuration of gravitating matter can exist only if the gravitational attraction is balanced by some repulsive forces~\cite{Volkov1999} which decrease~$r_h$ and ~$r_{\!\scriptscriptstyle{I\!\!\;SCO}}$. In our case, the radial pressure, $T_{11}={\phi'}^2f - 2V$ (in the orthonormal basis), will be positive and repulsive near the horizon if $V<0$ in some region around the horizon; this condition is necessary for the existence of a static scalar field configuration~\cite{Bekenstein1996}. Second, in the family of NSs, the metric functions $A(r,a)$ and effective potentials $\mathcal{V}_{\!e}(r,a,J)$ plotted in Fig.\,\ref{fig1} have minima for small and even zero values of $J$. For example, the curve $A(r,6.5)$ ($=\mathcal{V}_{\!e}(r,6.5,0$)) has a minimum at the point $r_m\approx0.79$. In fact, $r_m\rightarrow0$ as $a\rightarrow6+0$. However, the conclusion of the greatest importance is that the matter surrounding the NS is eventually captured by the gravitational potential well. Particles of gas or dust will cool down and be compressed by gravitational forces into a "gray shell" in the vicinity of the sphere of radius $r_m$.

\begin{figure}[]
\begin{center}\hspace{-2em}
\includegraphics [width=0.487\textwidth]{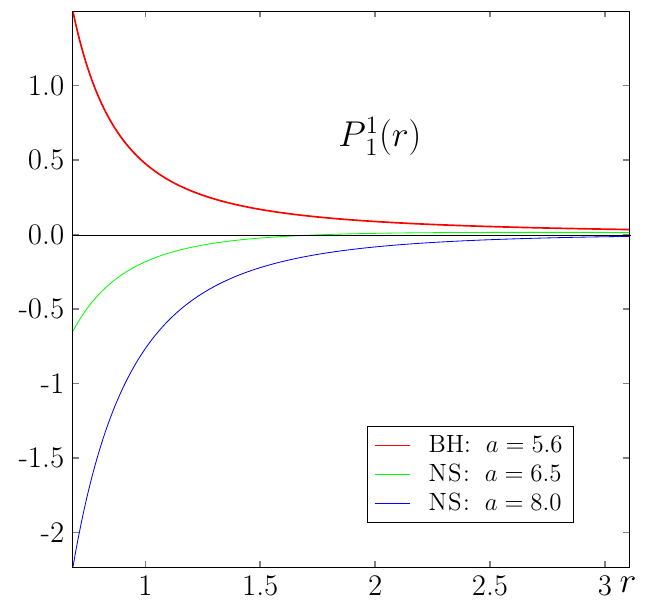}
\includegraphics [width=0.5\textwidth]{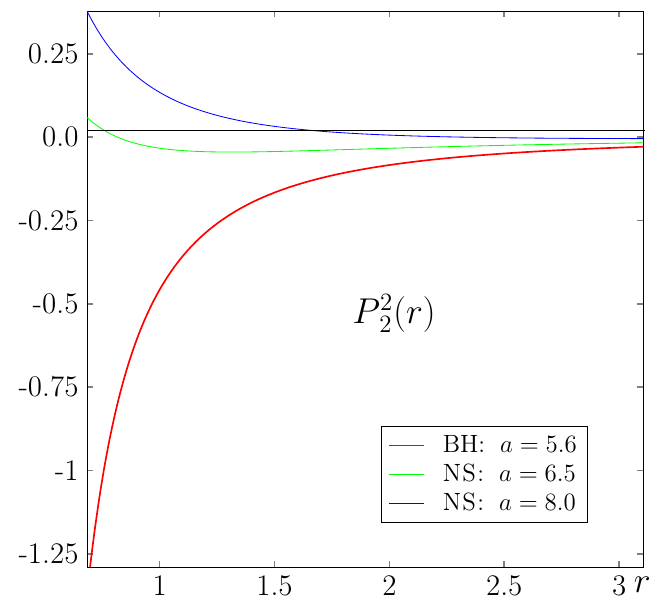}
\end{center}
\caption{The tidal tensor components are plotted here ($M=1$). The component $P_{\,3}^3$ is not presented, since it is approximately equal to $P_{\,2}^2$. The values of $a$ are the same as in~Fig.\,\ref{fig1}.}
\label{fig2}
\end{figure}

\section{Tidal Forces}
\label{Sec4}

The components of specific tidal force (acceleration) can be expressed as
\begin{equation}
F_t^{i}\equiv\frac{D^2\eta^{i}}{ds^2}= R^{i}_{jkl} U^{j}U^{k}\eta^{l} = P^{i}_{l}\eta^{l},\nonumber
\end{equation}
where $\eta^i$ is the coordinate distance between the center of a star and some point on the outer surface of the star, $U$ is its four-velocity in the coordinate frame, and $P=R(\cdot,U,U,\cdot)$ is the tidal tensor. For the metric~(\ref{metric}) with the metric functions given by the quadratures~(\ref{F-xi}) and~(\ref{A-f}), the components of the tidal tensor are obtained in~\cite{Andre2020}. In the rest frame (in the so-called instantaneous local inertial frame) of a star passing through the pericenter of its orbit, these components can be written explicitly as
\begin{equation}\label{P}
P= \left(\!1-\frac{rA'}{2A}\right)^{\!\!-1}\! \big[u(r)\varepsilon_1\otimes\varepsilon^1 + v(r) \varepsilon_2\otimes\varepsilon^2 + w(r) \varepsilon_3\otimes\varepsilon^3\big],
\end{equation}
where
\begin{equation}\label{Pcomp}
u(r)= \frac{1-f}{r^2}+ {\phi'}^2f+ \frac{f'A'}{4A}, \quad
v(r)= -\frac{f'}{2r}- {\phi'}^2f- \frac{A'}{2A}\frac{1-f}{r},
\quad
w(r)= -\frac{f'}{2r}- {\phi'}^2f
\end{equation}
and $\{\varepsilon_0=\varepsilon_t, \varepsilon_1=\varepsilon_r, \varepsilon_2=\varepsilon_\theta, \varepsilon_3=\varepsilon_\varphi\}$ is an orthonormal (with respect to the metric) basis  connected with the center of mass of the star.
It is easy to verify that for BHs the factor in front of the square brackets in~(\ref{P}) is always positive in the region in which $\xi>3M$, that is, outside the photon sphere (and everywhere for NSs). For a star falling radially, the components of the tidal force will be different, but we do not consider this case due to the negligible probability of such events.

In the pericenter, the radial instantaneous velocity is zero, and therefore the transition to the basis $\{\varepsilon\}$ is simple~\cite{Goel2015, Shahzad2017, Andre2020}. First, the orthonormal tetrad associated with the metric~(\ref{metric}) is defined by the formulae
\begin{equation}\label{e0}
{{\rm e}_{0}} = \frac{1}{\mathrm{e}^{F}\!\sqrt{f}\,}\,\partial_{t}, \quad
\mathrm{e}_{1} = \sqrt{f}\,\partial_{r},\quad
\mathrm{e}_{2} = \frac{1}{r}\,\partial_{\theta},\quad \mathrm{e}_{3} =
\frac{1}{r\sin\theta}\,\partial_{\varphi},
\end{equation}
\begin{equation}\label{e^0}
\mathrm{e}^0\!=\!\mathrm{e}^{F}\!\sqrt{f}\,dt,\quad \mathrm{e}^1 = \frac{1}{\sqrt{f}\,}\,dr,\quad \mathrm{e}^2 = r\,d\theta,\quad
\mathrm{e}^3
 = r\sin\theta\,d\varphi.
\end{equation}
Then the nonzero tetrad components of the four-velocity of a star, $U=\hat{U}^i{\rm e}_i$, can be expressed in terms of the integrals of motion~(\ref{int-motion}) as
$\hat{U}^0=E/\sqrt{A}$ and $\hat{U}^3=J/r$, so that the corresponding Lorentz transformation gives
\begin{equation}\label{eps0}
\varepsilon_0\!=\!U\!=\! \hat{U}^0{\rm e}_0+\hat{U}^3{\rm e}_3, \quad
\varepsilon_1\!=\!{\rm e}_1, \quad \varepsilon_2\!=\! {\rm e}_2, \quad
\varepsilon_3\!=\!\hat{U}^3{\rm e}_0+\hat{U}^0{\rm e}_3,
\end{equation}
\begin{equation}\label{eps^0}
  \varepsilon^0 = \hat{U}^0{\rm e}^0-\hat{U}^3{\rm e}^3, \quad
  \varepsilon^1\!=\! {\rm e}^1, \quad \varepsilon^2\!=\!{\rm e}^2, \quad
  \varepsilon^3\!=\!-\hat{U}^3{\rm e}^0+\hat{U}^0{\rm e}^3.
\end{equation}

The results of numerical calculations of the tidal tensor components are shown in~Fig.\,\ref{fig2}. The values of the parameter $a$ correspond to those in~Fig.\,\ref{fig1}. We have plotted only components $P_{\,1}^1$ and $P_{\,2}^2$, since $P_{\,3}^3\approx P_{\,2}^2$; this is not surprising because the system as a whole is invariant under the rotations around its radial axis of symmetry. First of all, it can be seen that a star approaching the central region of a BH will be stretched along the radial axis (component $P_{\,1}^1$ is positive), but it will undergoes tidal compression in the azimuthal directions ($P_{\,2}^2$ and $P_{\,3}^3$ are negative). In contrast, in the vicinity of a scalar field NS, a star will be compressed in the radial direction and stretched in the azimuthal directions. Thus, during the passage of the pericenter near the BH, a star has the shape of needle directed towards the center, while near a NS, it looks like a "pancake" oriented perpendicularly to the radial direction. It is also important that for the BH presented in~Fig.\,\ref{fig2}, the values of tidal forces near the ISCO are two orders of magnitude greater than those for the corresponding Schwarzschild BH.

\begin{figure}[]
\begin{center}\hspace{-2em}
\includegraphics [width=0.518\textwidth]{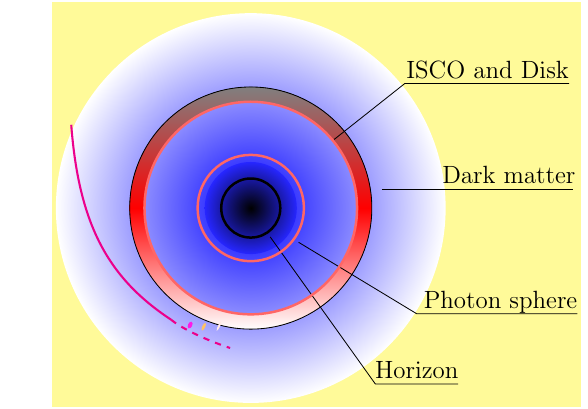}
\includegraphics [width=0.47\textwidth]{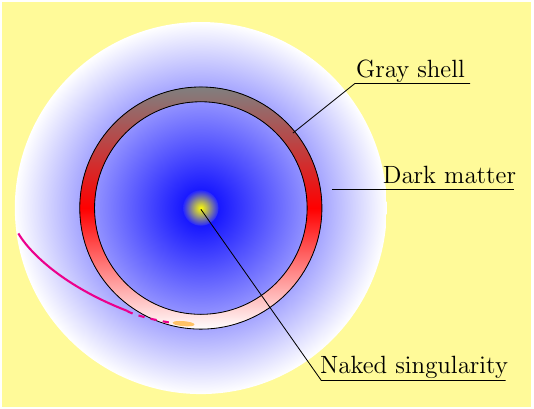}
\end{center}
\caption{Schematic comparison of the mechanisms of the emergence of flares in BHs and NSs.The left panel shows the tidal disruption of a star near the accretion disk of a BH. One part of the debris is captured by the BH and falls onto the disk, causing a flare, while the other part escapes from the pericenter. The right panel shows the collision of a star having a small angular momentum with the gray shell formed by baryonic matter in the potential well.}
\label{fig3}
\end{figure}

Hydrodynamic models of TDEs are quite complicated and crucially depend on the equation of state of the matter composing a star~\cite{Gezari2021, Banerjee2021, Lodato2015, Rossi2021}. Nevertheless, one can obtained a rough but useful estimate for the tidal radius $r_t$ from the condition that the disruption of the star begins if the attractive gravitational force on its surface becomes equal to the tidal force. That is, the outflow of matter begins when $M_*/R_*^2=P(r_t)R_*$, where $R_*$ and $M_*$ are, respectively, the radius and mass of the star, and $P(r_t)$ is the largest component of the tidal tensor; here we temporarily do not fix the unit of length by the condition $M=1$ (recall, however, that $G=c=1$ and therefore $[P]=cm^{-2}$). In the case of a vacuum BH, one has $P(r_t)=M/r_t^3$, and hence the tidal radius is determined by the well-known expression $r_t=R_*(M/M_*)^{1/3}$~\cite{Hill1975}. In the case of scalar field configurations, the tidal radii are determined implicitly as
\begin{equation}\label{}
P_1^1(r_t)=M_*/R_*^3 \;\;\text{and}\;\; P_2^2(r_t)=M_*/R_*^3
\end{equation}
for BHs and NSs, respectively, where the components
\begin{equation}\label{}
P_1^1= \left(\!1-\frac{rA'}{2A}\right)^{\!\!-1}\! \left(\frac{1-f}{r^2}+ {\phi'}^2f+ \frac{f'A'}{4A}\right), \quad
\end{equation}
\begin{equation}\label{}
P_2^2= -\left(\!1-\frac{rA'}{2A}\right)^{\!\!-1}\! \left(\frac{f'}{2r}+ {\phi'}^2f+ \frac{A'}{2A}\frac{1-f}{r}\right)
\end{equation}
are defined in accordance with the expressions~(\ref{P}) and~(\ref{Pcomp}).

Suppose that a solar-like star with $M_*=1.5\cdot10^{5}cm\approx{}M_\odot$ and $R_*=10^{10}cm$ (core radius) moves near a supermassive BH (for example, Sgr A*) with mass $M\approx6\cdot10^{11}cm$. A numerical calculation gives $r_t\approx2.65M$ for both the vacuum BH, and the scalar field BH presented in~Fig.\,\ref{fig1} with $a=5.6$. The vacuum BH will capture the star without its tidal disruption, while in the second case, the star will begin to destroy at a distance of more than 8.5 horizon radii ($r_h\approx0.31$). This example shows the influence of DM on TDEs. Note that the same effect takes place for scalar field NSs.

\section{Discussion}\label{Disscussion}

Our current understanding of the TDEs is based primarily on numerical simulations, which in turn are based on very strong assumptions and various approximations. In particular, we also do not fully understand how the accretion disk generates a bright flare, that is, do not understand the mechanism that converts the kinetic energy of the debris of a destroyed and subsequently accreted star into radiation~\cite{Banerjee2021, Charalampopoulos2022, Wevers2022}. Both analytical and numerical models of the process of stellar tidal disruption \textit{in the vicinity of a supermassive BH} are presented in works~\cite{Lodato2015, Rossi2021, Lodato2011, Liu2021}, as well as in the literature cited there. In our paper we have tried to consider in detail two circumstances related to TDEs. First, the centers of galaxies are surrounded by dark matter, so that the gravitational attraction acting on a star decreases with decreasing radius of its orbit. It turns out, that the event horizon radius of a scalar field BH can be arbitrarily small as well as the ISCO radius, and tidal forces near the ISCO can be very large. Second, if strongly gravitating objects in the centers of galaxies are NSs, then the mechanism of the emergence of bright flares can be completely different. In the vicinity of a scalar field NS, baryonic matter is concentrated in the gravitational potential well forming a "gray shell" consisting of cold compressed matter. In this case, the flares are caused by the collision of stars with this shell (see Fig.\,\ref{fig3}). We believe that future astrophysical observations of TDEs will clarify the question of which of these mechanisms is correct and will help to understand the geometric structure of spacetime near the centers of galaxies.


\end{document}